\documentclass[conference]{IEEEtran}

\ifCLASSINFOpdf
   \usepackage[pdftex]{graphicx}
 \else
   \usepackage[dvips]{graphicx}
 \fi
\usepackage{algorithmic}

\usepackage{multirow}
\usepackage{amsmath, amssymb}
\usepackage{array}
\usepackage{subfigure}
\usepackage{amsmath}
\usepackage{pseudocode}

\begin{document}
\title{Sampling from Diffusion Networks}

\author{\IEEEauthorblockN{Motahareh Eslami Mehdiabadi\IEEEauthorrefmark{1}, Hamid R. Rabiee\IEEEauthorrefmark{2} and Mostafa Salehi\IEEEauthorrefmark{1}}
Department of Computer Engineering, Sharif University of Technology\\
\IEEEauthorblockA{\IEEEauthorrefmark{1}
Email: \{eslami, mostafa$\_$salehi\}@ce.sharif.edu} \IEEEauthorblockA{\IEEEauthorrefmark{2}
Email: rabiee@sharif.edu}}

\maketitle
\begin{abstract}
The diffusion phenomenon has a remarkable impact on Online Social Networks (OSNs). Gathering diffusion data over these large networks encounters many challenges which can be alleviated by adopting a suitable sampling approach. 
The contributions of this paper is twofold. First we study the sampling approaches over diffusion networks, and for the first time, classify these approaches into two categories; (1) Structure-based Sampling (\textsc{Sbs}), and (2) Diffusion-based Sampling (\textsc{Dbs}). 
The dependency of the former approach to topological features of the network, and unavailability of real diffusion paths in the latter, converts the problem of choosing an appropriate sampling approach to a trade-off.
Second, we formally define the diffusion network sampling problem and propose a number of new diffusion-based characteristics to evaluate introduced sampling approaches.
Our experiments on large scale synthetic and real datasets show that although \textsc{Dbs} performs much better than \textsc{Sbs} in higher sampling rates ($16\%\sim29\%$ on average), their performances differ about $7\%$ in lower sampling rates. Therefore, in real large scale systems with low sampling rate requirements, \textsc{Sbs} would be a better choice according to its lower time complexity in gathering data compared to \textsc{Dbs}. Moreover, we show that the introduced sampling approaches (\textsc{Sbs} and \textsc{Dbs}) play a more important role than the graph exploration techniques such as Breadth-First Search (BFS) and Random Walk (RW) in the analysis of diffusion processes.

\end{abstract}

\section{Introduction}

\begin{figure*}[htp]
  \begin{center}
    \subfigure[Structure-based Sampling (\textsc{Sbs})]{\label{structure}\includegraphics[scale=0.1]{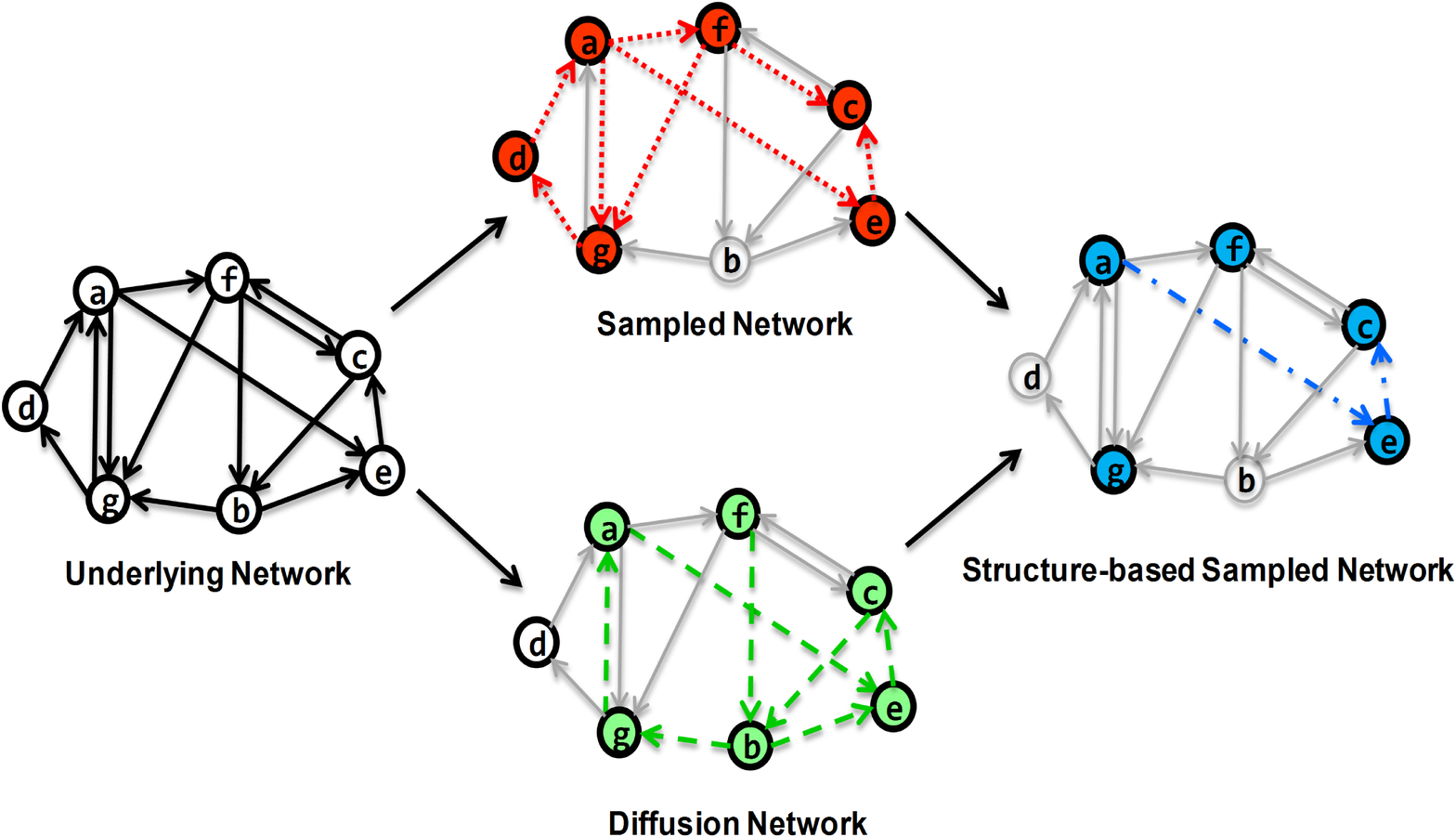}}
    \subfigure[Diffusion-based Sampling\textsc{Dbs}] {\label{diffusion}\includegraphics[scale=0.1]{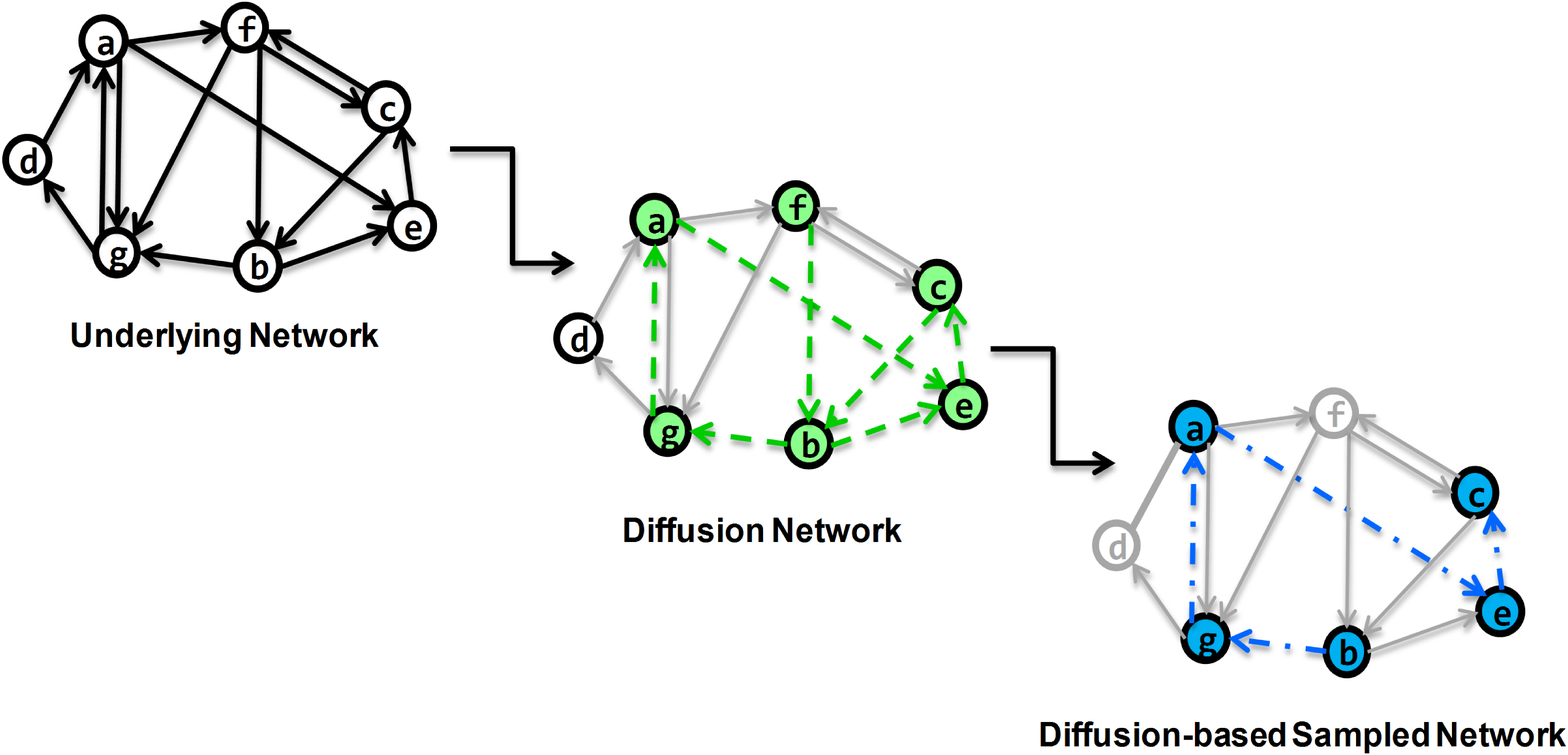}}
 \end{center}
  \caption{
  \small{Sampling Approaches. The dashed, dotted and dash-dot lines illustrate the links of diffusion, sampled and the final generated networks, respectively. 
  } 
  } 
  \label{approaches}
\end{figure*}

Information diffusion as a new area of multi-disciplinary research has a remarkable effect on social networks \cite{Gomez10}. 
In recent years, large Online Social Networks (OSN) such as Facebook, Twitter and YouTube have been the source of information propagation in different formats such as posts, tweets, and videos. The growth in the size of these networks results in large information networks. For example, in March 2011, Twitter users were sending 50 million tweets per day \cite{twitter}.
Therefore, collecting the diffusion data over large scale OSNs is often infeasible in many applications. This challenge necessitates the need for manipulating the diffusion data in an efficient way to analyze the diffusion process behavior. 

Sampling strategy can be considered as a solution to solve this problem by decreasing the expense of processing on large real networks. In recent years, a considerable amount of research has been done on analyzing the topological characteristics of large OSNs based on the sampled data of different networks such as Facebook \cite{gjoka2011}, Twitter \cite{Salehi2011}, YouTube \cite{mislove2007}, and other large networks \cite{Leskovec2006, Salehi12}. However, considering the sampling approaches to study diffusion behaviors of social networks, apart from their topologies, is a remarkable issue that should be addressed. 

To the best of our knowledge, there is no comprehensive study about sampling approaches on diffusion process. Looking into several large diffusion network studies \cite{ Nowell08, Choudhury10, Sadikov11, Gomez11}, we classify the data gathering approaches of diffusion process into two categories; (1) Structure-based Sampling (\textsc{Sbs}), and (2) Diffusion-based Sampling (\textsc{Dbs}). The former approach is based on the topology of the network and the latter considers propagation paths and diffusion process in sampling methodology. The \textsc{Sbs} approach will produce some redundant data that results in decreasing the accuracy of diffusion process measurements. On the other hand, The\textsc{Dbs} approach can reduce the redundant data and consequently increase the sampling efficiency by following the diffusion paths. However, obtaining the real diffusion paths is not practical in many applications \cite{Gomez10, Eslami11}. These challenges converts the problem of choosing an appropriate sampling approach for diffusion process analysis to a trade-off between many parameters such as the amount of the sampled data and the availability of diffusion paths.

In this paper, we evaluate the performance of the proposed sampling approaches to show how different sampling approaches can impact the measurement of diffusion process. To this end, our methodology comprises two steps. First, we formally define the diffusion network sampling problem. Second, we propose a number of new evaluation characteristics for the diffusion process in order to analyze their behaviors based on different sampling approaches. The proposed characteristics alleviate the dependency to topological features of the network and increase the correlation with the diffusion process. Moreover, we categorize these characteristics into three classes; (1) node-based, (2) link-based, and (3) cascade-based.

We analyze the introduced approaches by extensive experiments over large synthetic and real datasets. Two well-known sampling techniques of Breadth-First Search (BFS) and Random Walk (RW) are used in both \textsc{Dbs} and \textsc{Sbs} approaches. Our experiments show that the accuracy of measuring node-based and link-based characteristics in \textsc{Dbs} grows more than \textsc{Sbs} by increasing the sampling rate. This phenomenon will result in a considerable performance difference between these approaches in higher sampling rates (up to 65\% difference). Nevertheless, cascade-based characteristics can decrease this performance difference compared to the node-based and link-based characteristics. This is the result of inherent difference between these characteristics as the formers are individual-based characteristics while the latter is related to the cascades as a group-based characteristic.

Our evaluation on performance between the proposed sampling approaches shows that \textsc{Sbs} is similar to \textsc{Dbs} in low sampling rates (the difference is about  $7\%$ in average). The results demonstrate that \textsc{Sbs} can be used in real systems in which only low sampling rates are feasible. Furthermore, we investigate the performance of RW and BFS in measuring diffusion characteristics. Our experiments reveal that these sampling techniques perform similar to each other (with $3\%$ difference in average). This clarifies that the proposed sampling approaches (\textsc{Dbs} and \textsc{Sbs}) have more important effect than the graph exploring techniques such as RW and BFS. 

In summary, our main contributions can be summarized in the following:
\begin{itemize}
\item
Classifying and proposing sampling approaches of \textsc{Sbs} and \textsc{Dbs} for diffusion process analysis
\item
Defining the diffusion network sampling formally
\item
Proposing and categorizing some new diffusion-based characteristics to study the behavior of sampling approaches
\item
Evaluating sampling approaches (\textsc{Sbs} and \textsc{Dbs}) and sampling techniques (BFS and RW) in different sampling rates to evaluate their performances
\end{itemize}

The rest of the paper is organized as follows. Section \ref{sec:related work} presents a classification of data collection approaches in the field of information diffusion networks. The problem definition is proposed in Section \ref{sec:Problem Formulation}. Section \ref{Experimental Evaluation} provides the performance evaluations, and the concluding remarks are provided in Section \ref{Conclusions}.

\section{Diffusion Network Data Collection}\label{sec:related work}

Diffusion networks have attracted considerable attention in recent years \cite{Gomez10, Nowell08,  Gomez11, Eslami11}. In spite of this great attention, there is no comprehensive survey on how to collect data from a diffusion network. The most close work to ours is \cite{Choudhury10}, which studies the impact of some sampling techniques (such as Random sampling) on the information diffusion process. This work does not consider sampling approaches and their effect on diffusion process analysis which we address in this paper. In the following, we propose a new classification of diffusion data sampling approaches.

\textbf{Structure-based Sampling:} The most common approach for sampling the diffusion process is to sample the underlying network and then extracting the diffusion links from the collected data (refer Figure \ref{structure}). Since this approach is based on the structure of the underlying network, and not the diffusion process, we call it the structure-based sampling approach (\textsc{Sbs}).
Sampling the Twitter network to study on the resulting diffusion network \cite{Choudhury10, Sadikov11} and inferring diffusion topics from the DBLP database \cite{Lin11} are some examples which utilize \textsc{Sbs} to analyze the diffusion process. Using this approach will result in extraction of some redundant data such as nodes and links which do not participate in the diffusion process. Therefore, these data should be removed from the collected data to obtain the sampled diffusion network. This data reduction leads to a smaller sampled graph which may decrease the accuracy of analysis. 

\textbf{Diffusion-based Sampling}: To study on diffusion networks, one may track the diffusion paths instead of the network paths. This idea leads to another sampling approach that explicitly considers the diffusion characteristics. We call this the diffusion-based sampling (\textsc{Dbs}) (refer to Figure \ref{diffusion}). Recently, this approach is used in \cite{Gomez10, Gomez11} to collect the diffusion data. 

Since diffusion network is a sub-graph of the underlying network, using \textsc{Dbs} will increase the accuracy of the diffusion process analysis. Moreover, this approach reduces the cost of data collection by sampling only the diffusion data (i.e. the redundant data is not collected). 
For example, comparing \textsc{Sbs} and \textsc{Dbs} in Figure \ref{approaches}, it is observable that with the same sampling rate of $0.5$ with respect to the edges, the resultant graph in \textsc{Dbs} approach contains more links which participate in the diffusion process than the one in \textsc{Sbs}. 
Nevertheless, \textsc{Dbs} can not be used in for applications in which direct access to the links of the diffusion network is not feasible. 

Actually, the latent nature of the diffusion network structure does not allow us to explore it (as simple as the underlying network) \cite{Gomez10, Eslami11}. Therefore, choosing an appropriate sampling approach will be a trade-off between many conditions such as the amount of the sampled data and the availability of diffusion paths. Recently, we have proposed a novel sampling technique by utilizing the diffusion process characteristics \cite{Eslami12}. In particular, it uses the infection times as local information without any knowledge about the latent structure of diffusion network.

\textbf{Sampling Techniques:}
Since the structure of the original network is unknown initially, we use the graph exploration techniques of Breadth-First Search (BFS) and Random walk (RW) in both \textsc{Sbs} and \textsc{Dbs} approaches.  BFS is a basic graph-based sampling technique that has been used extensively for sampling the networks in various domains \cite{Salehi2011, mislove2007,  Wilson09}. At each iteration of BFS, the earliest explored node is selected next, and eventually, all nodes within some distance from the starting node is discovered.
RW \cite{Lovas93} is also one of the most widely used exploration sampling techniques in different kind of network contexts such as uniformly sampling Web pages from the Internet \cite{henzinger2000}, degree distributions of the Facebook social graph \cite{gjoka2011} and in general large graphs \cite{Leskovec2006}. A classic RW samples a graph by moving from a node $u$, to a neighboring node $v$, through an outgoing link $(u,v)$, chosen uniformly at random from the neighbors of node $u$. 

\section{Diffusion Network Sampling}\label{sec:Problem Formulation}
\subsection{Preliminaries}
Let $G = (V, E)$ with $n=|V|$, and $m=|E|$ be the graph representing a social network, where $V$ is the set of nodes, and $E$ is the set of unweighted links between pairs of nodes. Network $G$ is called the underlying network since the information diffusion process will occur over $G$. 
Spreading some diffusible chunks of information over the underlying network creates a path which is called a cascade. A cascade can transmit some pieces of information such as epidemic diseases; Therefore, we may refer to these diffusible chunks as ``infection" \cite{Eslami11}. Each cascade $c$ has $n_{c}$ edges that is shown by an Infection Vector (IV) in which the order of edges illustrates the order of cascade passage over them:
\begin{equation}
 IV_{c} = {\{e_{1},e_{2},\cdots, e_{n_{c}}\}}
\end{equation}

The transmission model of cascades in this work follows the independent cascade model of \cite{Gomez10}. In this model, each node infects each of its neighbors independently by a random variable. Propagating these information cascades over the underlying network builds the diffusion network which is called $G^{*} = (V ^{*}, E^{*})$.  The covering percentage of diffusion network over the underlying network depends on a metric, called the diffusion rate $\delta$. 
The parameter that controls how far a cascade can spread is denoted by $\beta$ \cite{Gomez10}.

We call an ``element set", $T$, which refers to a set of diffusion network elements that could be nodes, links or cascades. For element $e \in T$, $L$ is defined as a finite set of labels which shows a specific feature of $e$. We assume that the label $l_e \in L$ is assigned to each element $e$  by a function $f:T \rightarrow L$ which is called the measurement function. For example, infection is a label for each node that shows whether this node is infected during the diffusion process or not. The measurement function $f$ for this label will match nodes $u \in V$ to the set  $L=\{0, 1\}$ ($f(u) = 0$, if node $u$ is not infected and $f(u) = 1$, otherwise).
To measure a characteristic of network $G$, 
We consider the average function $A_{G}(f)$ that is defined as: 

\begin{equation} \label{probDef.}
A_{G}(f)=\frac{\sum_{u\in V} {f(u)} }{|V|}
\end{equation}

In the infection example, this average shows the percentage of infected nodes by the diffusion process to all the nodes of the underlying network.

\subsection{Problem Definition} \label{problemDef.}

Consider the graph $G$ as the underlying graph for a sampling approach that will yield the sampled graph $G(S)$ as a sub-graph of $G$. Let define the accuracy of a sampling approach as: 

\begin{equation}
\lambda = 1 - \frac{{|A_{G^{*}}(f) - A_{G(S)}(f)| }} {A_{G^{*}}(f)}
\end{equation}

Our goal is to evaluate the proposed sampling approaches (\textsc{Sbs} and \textsc{Dbs}), in terms of the accuracy in measuring the characteristics of diffusion process. 
The sampling rate  $\mu$ and diffusion rate $\delta$ are the constraints of this problem. 

\begin{figure*}[t]
  \begin{center}
  %************************SEED************************
    \subfigure[\small{Core Periphery Network}]{\includegraphics[scale=0.075]{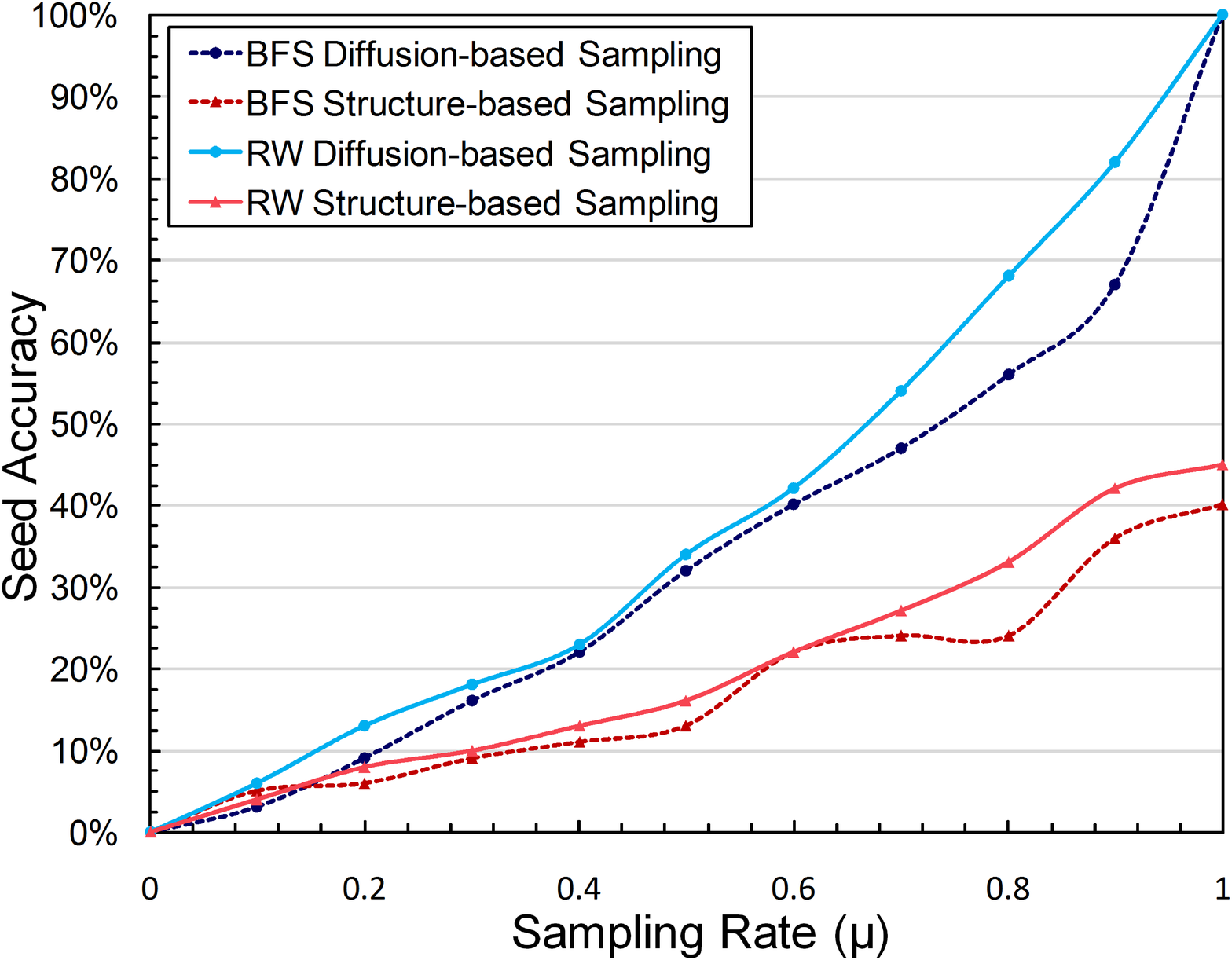}}
    \subfigure[\small{Hierarchical Network}] {\includegraphics[scale=0.075]{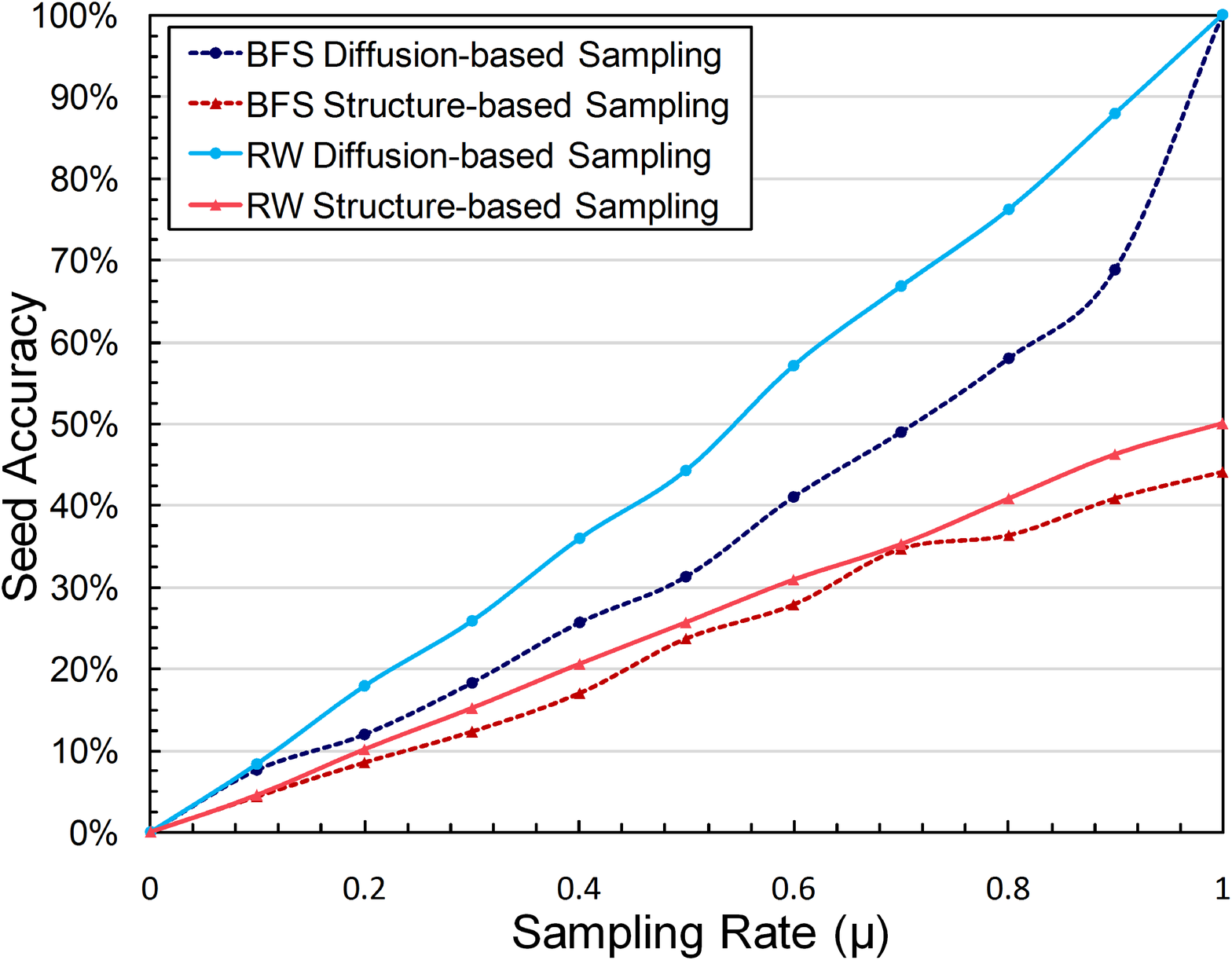}}
     \subfigure[\small{Random Network}] {\includegraphics[scale=0.075]{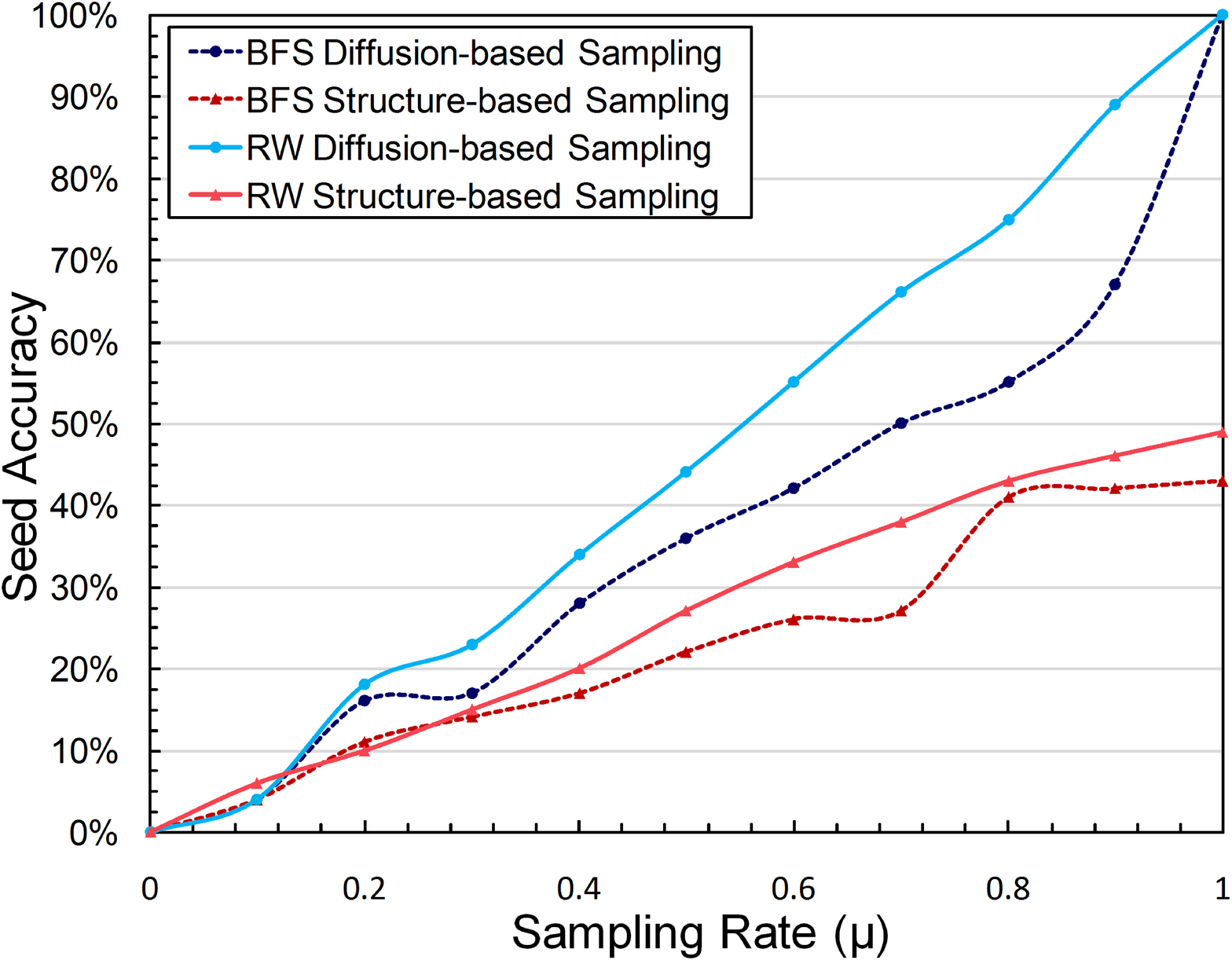}}
     \subfigure[\small{Forest Fire Network}]{\includegraphics[scale=0.075]{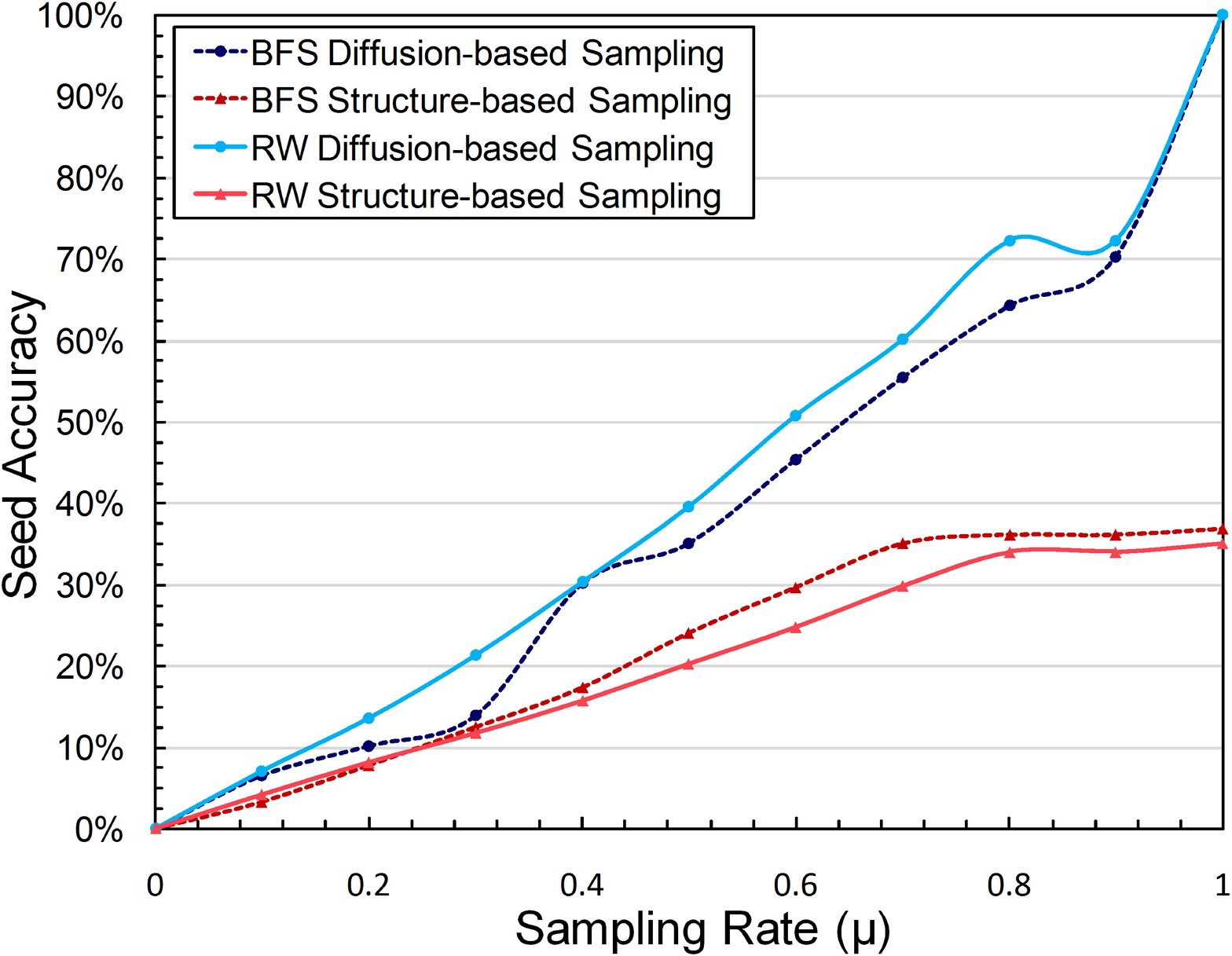}}
     \subfigure[\small{Political Blogsphere Network}] {\label{blog-seed}\includegraphics[scale=0.075]{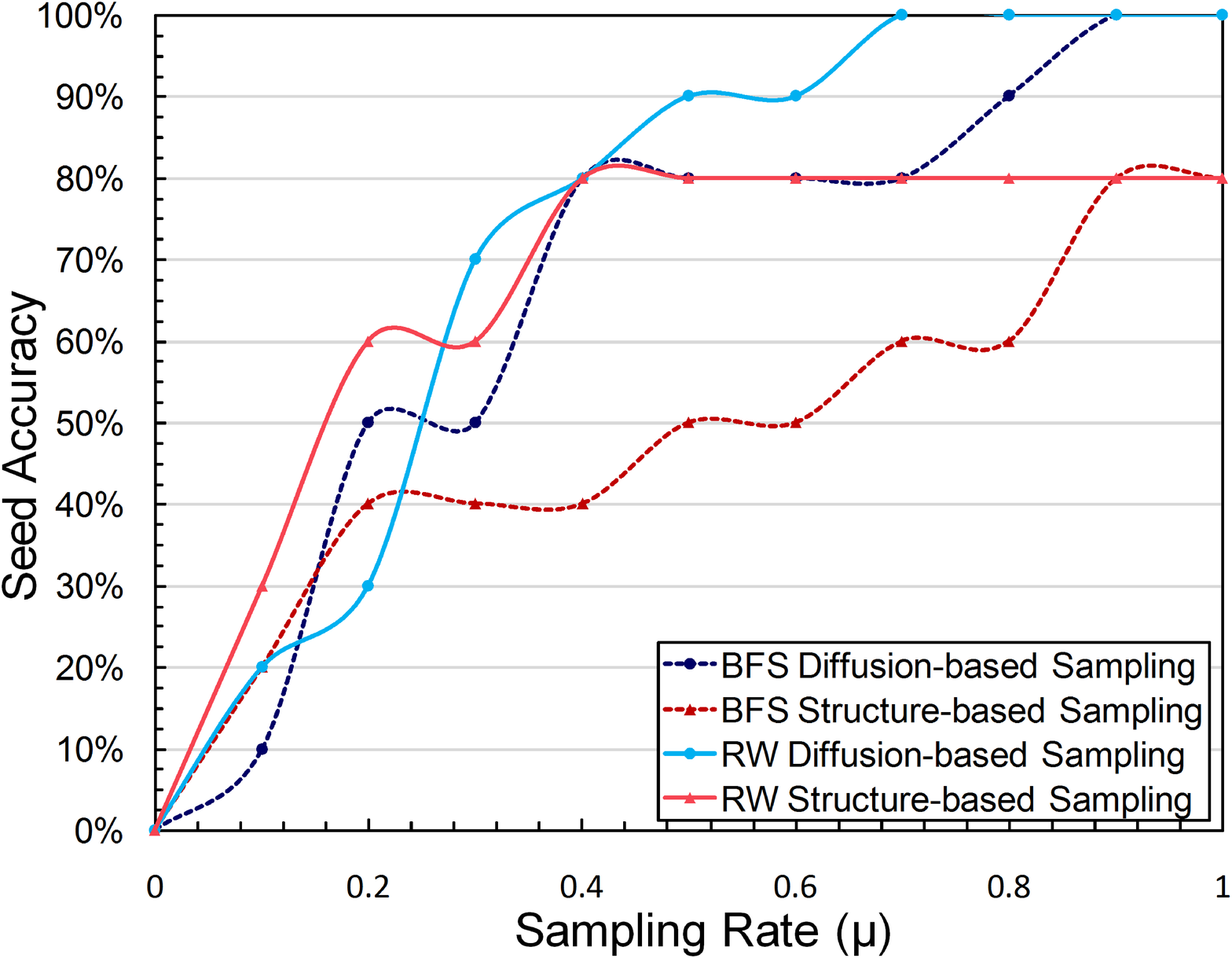}} 
     \subfigure[\small{Co-authorship Network }]{\includegraphics[scale=0.075]{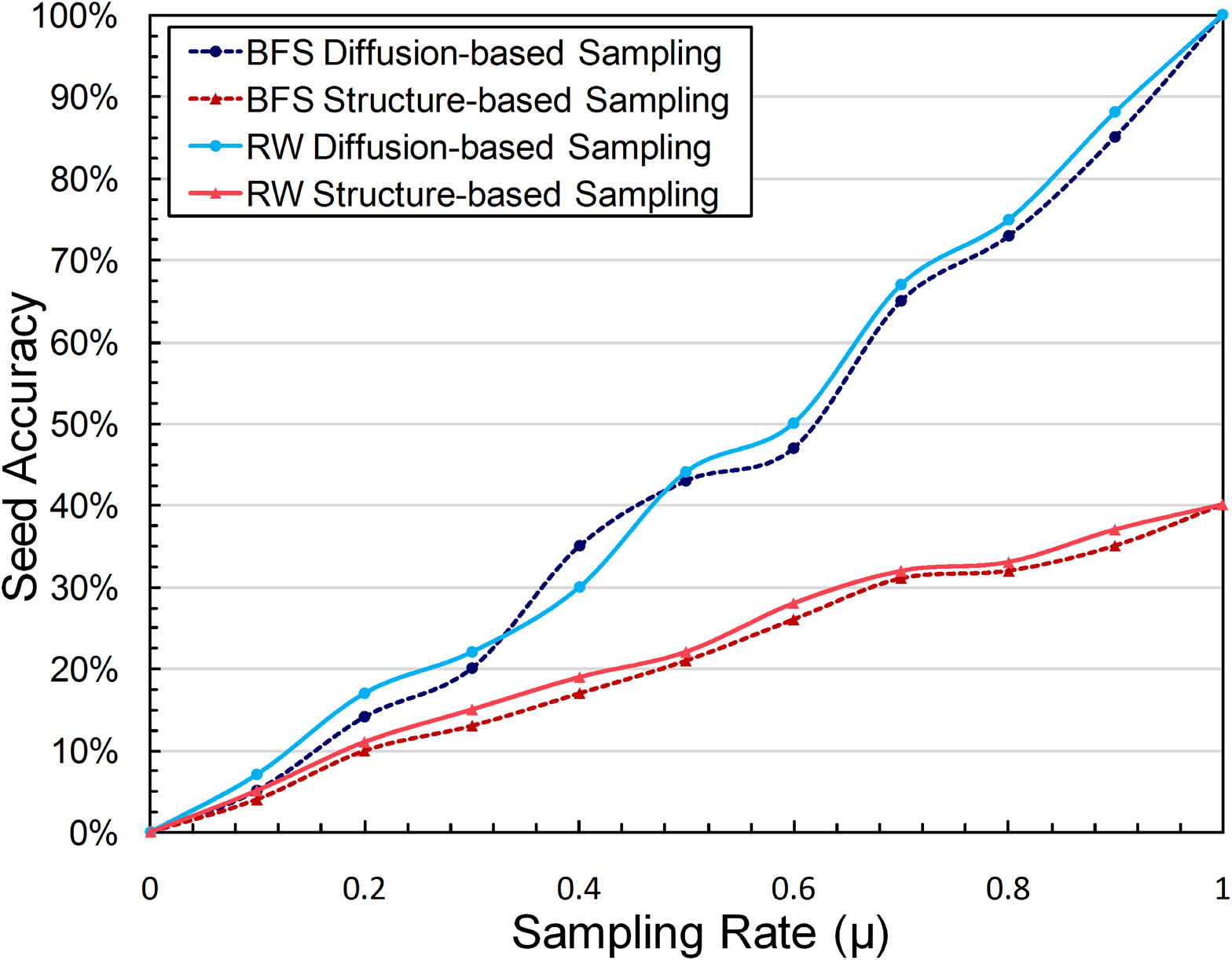}}
  \end{center}
  \caption{\small{Accuracy of seed characteristic measurement in different sampling approaches.}}
  \label{Seed}
\end{figure*}

\begin{figure*}[t]
  \begin{center}
  %************************SEED************************
    \subfigure[\small{Core Periphery Network}]{\includegraphics[scale=0.075]{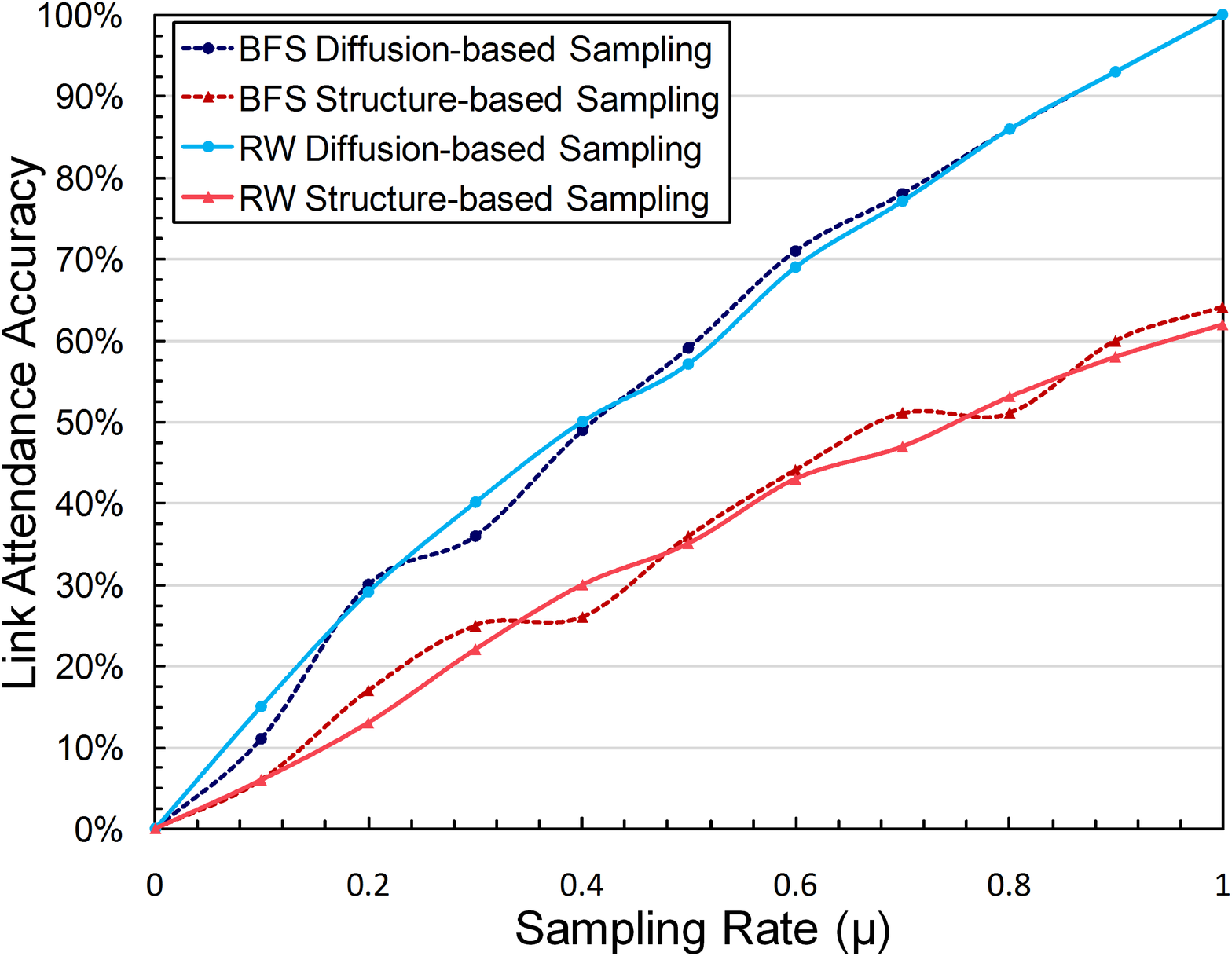}}
     \subfigure[\small{Hierarchical Network}] {\includegraphics[scale=0.075]{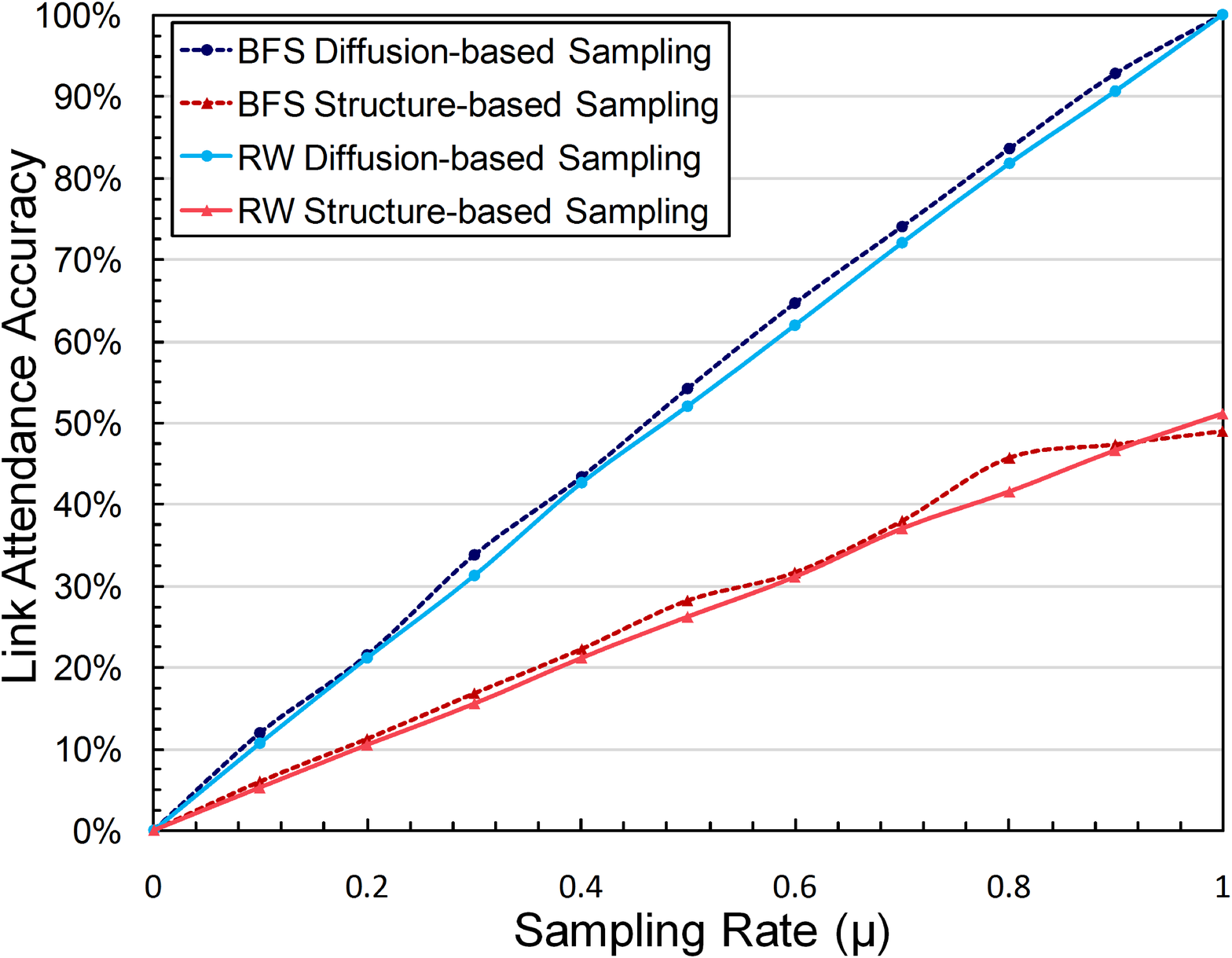}}
     \subfigure[\small{Random Network}] {\includegraphics[scale=0.075]{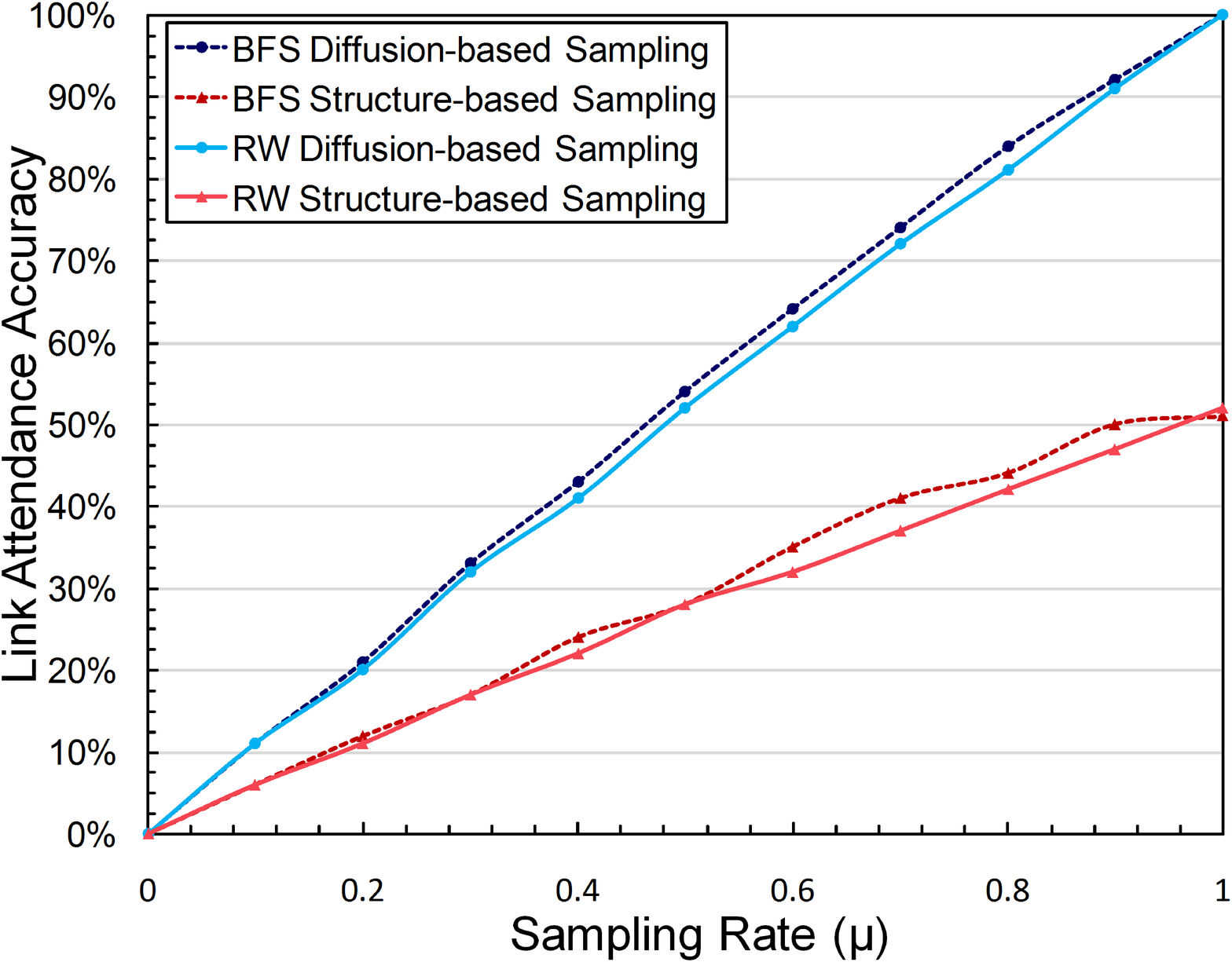}}
         \subfigure[\small{Forest Fire Network}]{\includegraphics[scale=0.075]{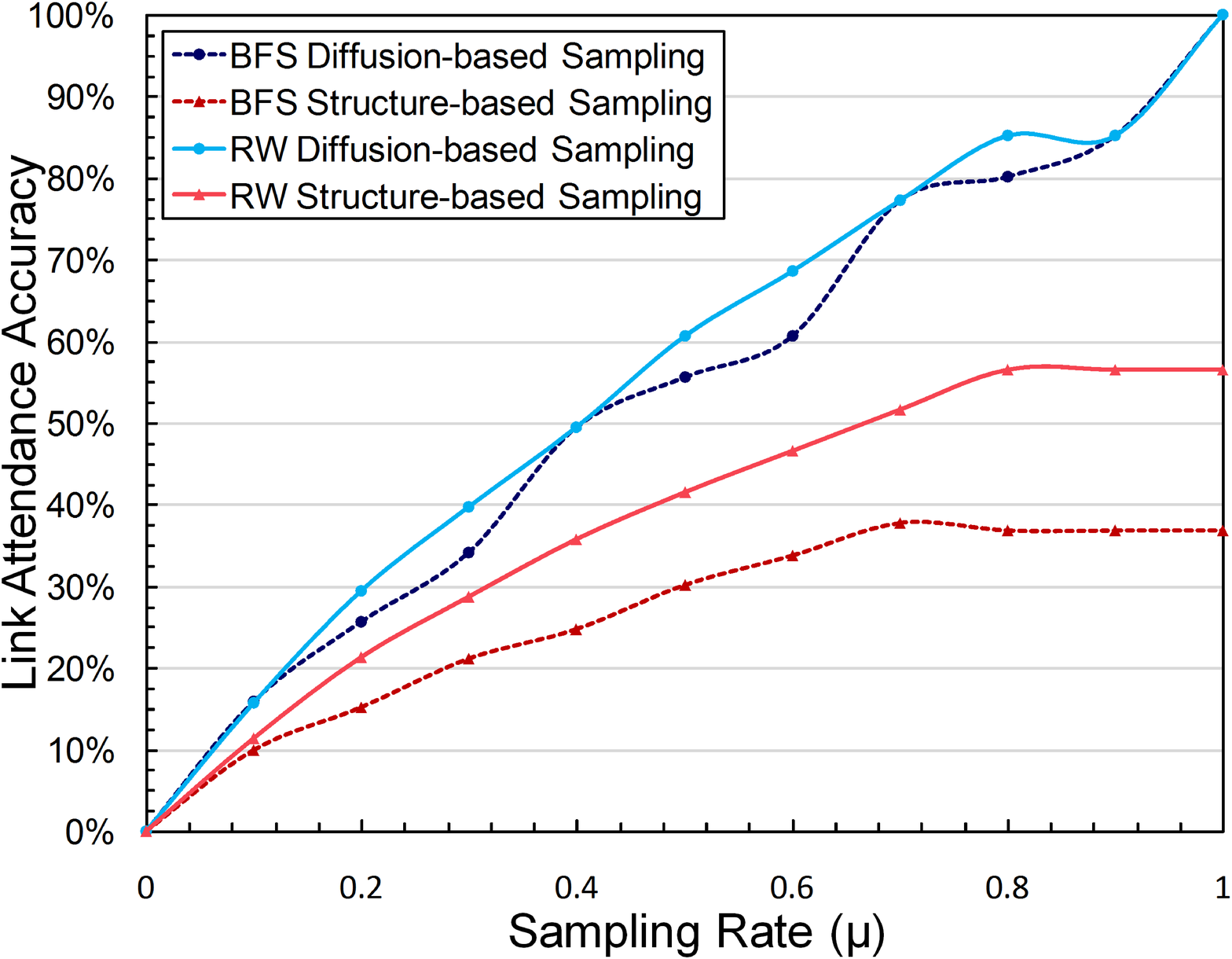}}
     \subfigure[\small{Political Blogsphere Network}] {\includegraphics[scale=0.075]{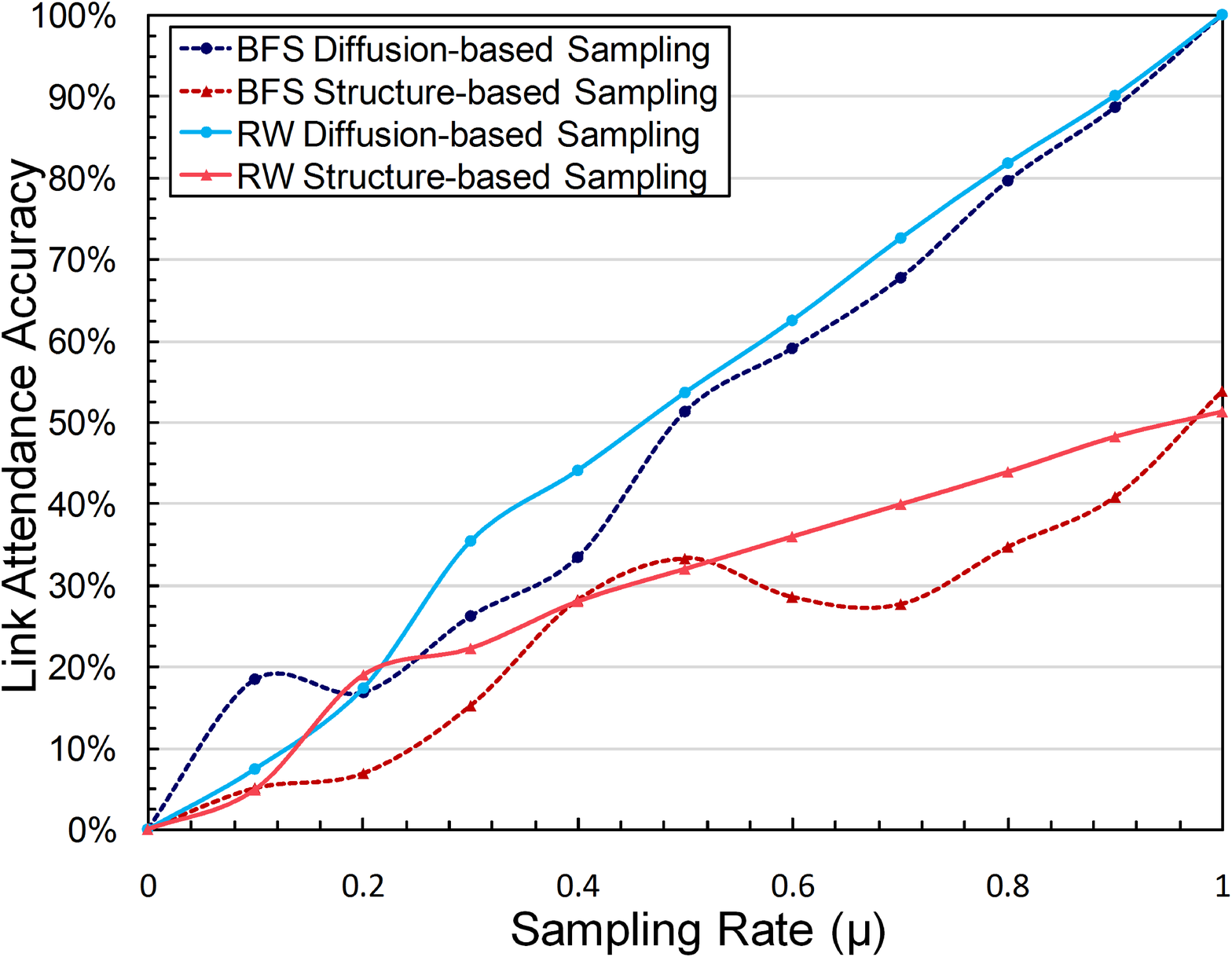}}
     \subfigure[\small{Co-authorship Network}]{\includegraphics[scale=0.075]{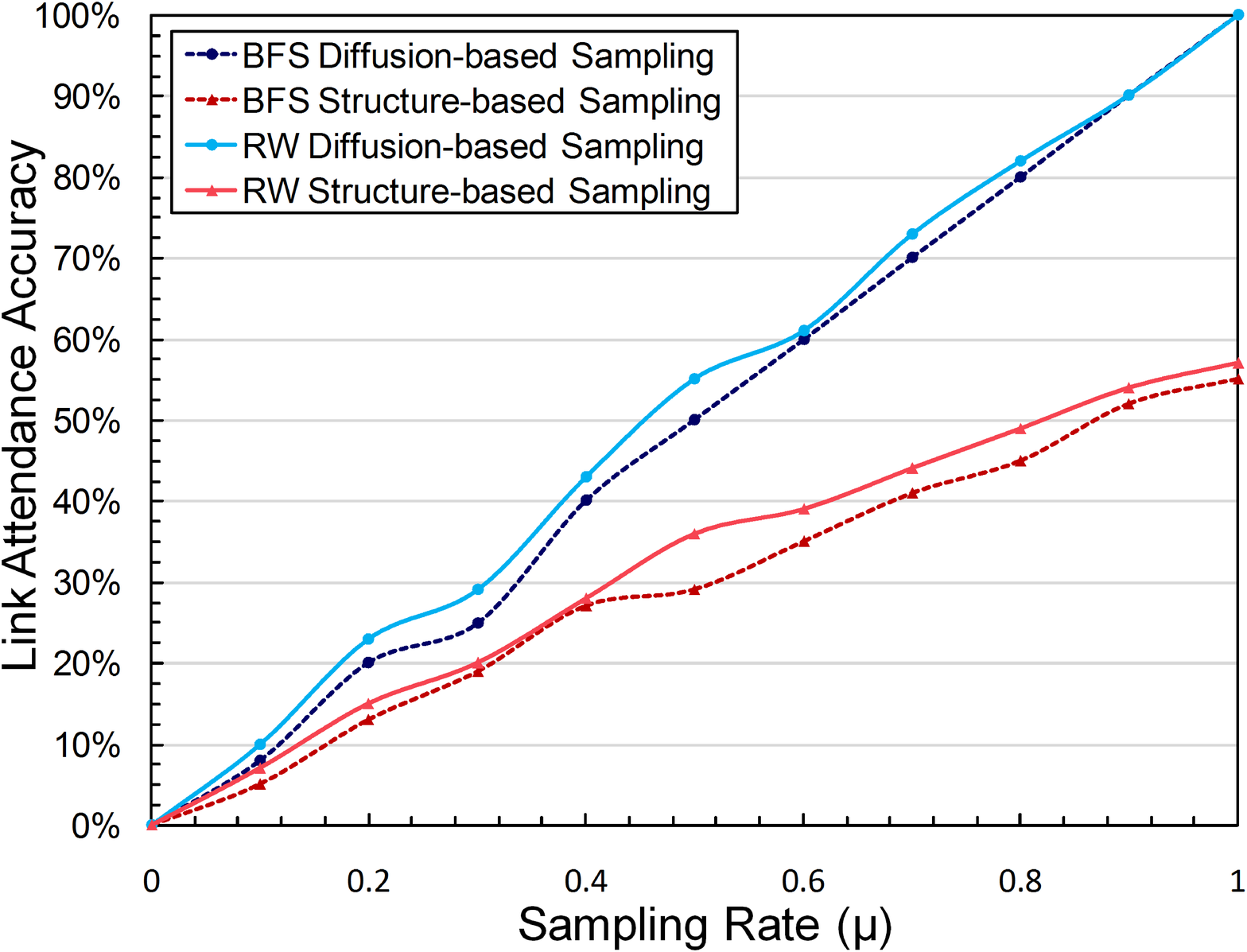}} 
  \end{center}
  \caption{\small{Accuracy of Link Attendance characteristic measurement in different sampling approaches.}}
  \label{Link}
\end{figure*}

\section{Experimental Evaluation}\label{Experimental Evaluation}

In this section, the performance of \textsc{Sbs} and \textsc{Dbs} will be analyzed by measuring a number of newly defined diffusion process characteristics. In both approaches, we use two sampling techniques of BFS and RW.  As \textsc{Sbs} and \textsc{Dbs} will be done over the underlying network and diffusion network respectively, we should consider different sampling ratios for each of these approaches to result in the same number of edges in the sampled network. Since diffusion network is a subset of the underlying network (by proportion of $\delta$), different sampling rates ($\mu$) from 0 to 1 for \textsc{Dbs} over diffusion network will be equal to $0<\mu<\delta$ for \textsc{Sbs} over the underlying network. For easier readability, we consider the sampling rate related to \textsc{Dbs} in all figures. For a reasonable rate of information diffusion, we also consider $\delta = 0.5$ which means $G^{*}$ will cover half of the $G$.

\subsection{Synthetic Dataset}
In order to construct synthetic networks, two well-known models for generating such networks are used: Kronecker \cite{LeskovecF07} and Forest Fire model \cite{LeKlFa05}. Using different sets of parameters in the Kronecker model, we generate three different networks named Random \cite{Erdos60}, Hierarchical \cite{Clauset08}, and Core-Periphery network \cite{LeskovecLDM08}. The parameters for generating networks and propagating cascades are provided in Table \ref{Parameters}.

\subsection{Real Dataset}
We use two real-world networks. The first network is a political blogosphere which has 1490 blogs and 19090 directed links between them \cite{Adamic05}.
The other is a co-authorship network of theory scientists that contains 2742 directed links between 1589 scientists \cite{Newman06}.

\begin{table}[]
\small
\caption{\small{The Setting Parameters.} 
\label{Parameters}}

\begin{center}
\begin{tabular}[c]{|l|c|c|c|c|c|} 
\hline
{ \textbf{Network} } & {\textbf{Parameter Matrix}} &  { \textbf{Nodes}} & {\textbf{Edges}} & { \textbf{ $\beta$}}\\

\hline
Core-Periphery & $[0.9, 0.5; 0.5, 0.3]$  & $8192$ & $15000$  & $0.1$  \\
\hline
Hierarchical  & $[0.5, 0.5; 0.5, 0.5]$ & $8192$ & $11707$  & $0.5$  \\ 
\hline
Random(ER) & $[0.9, 0.1; 0.1, 0.9]$  & $8192$ & $15000$  & $0.5$  \\
\hline
Forest Fire & $[5;0.12;0.1;1;0]$  &$10000$ & $14305$  & $0.5$ \\ 
\hline
\end{tabular}
\end{center} 
\end{table}

\subsection{Diffusion Characteristics Evaluation} 

\begin{figure*}[t]
  \begin{center}
    \subfigure[Core Periphery Network]{\includegraphics[scale=0.073]{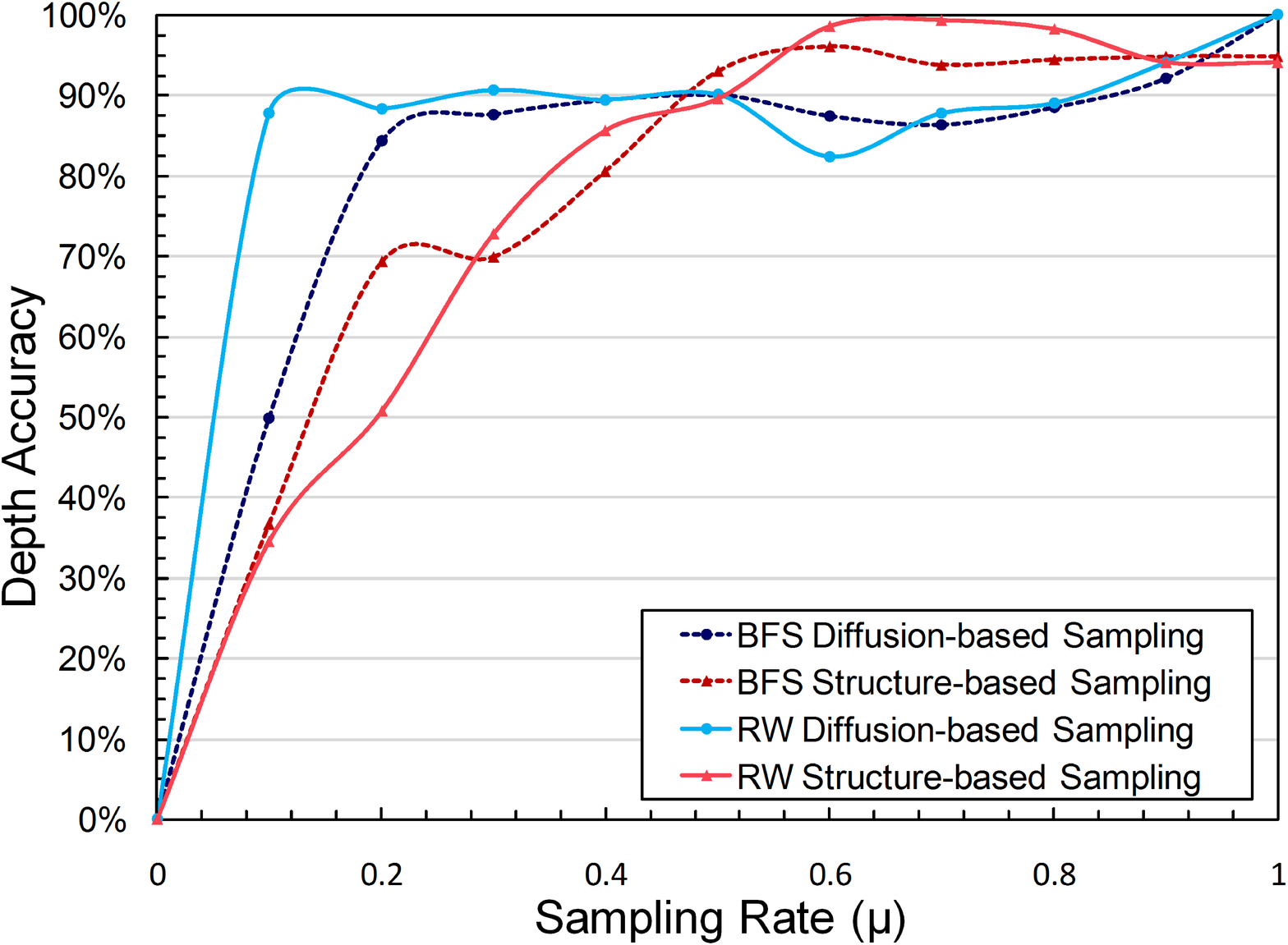}}
    \subfigure[Hierarchical Network] {\includegraphics[scale=0.073]{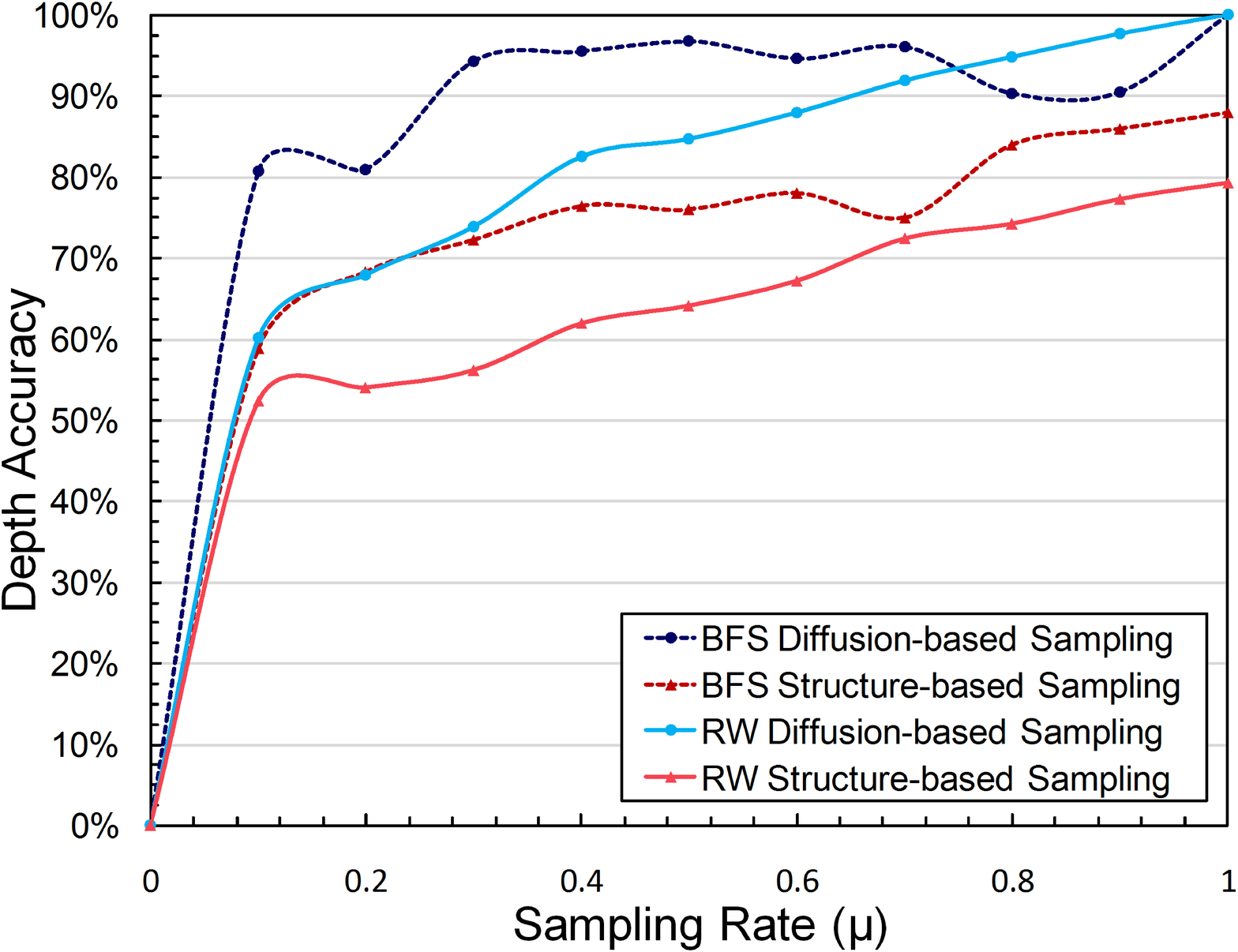}}
    \subfigure[Random Network] {\includegraphics[scale=0.073]{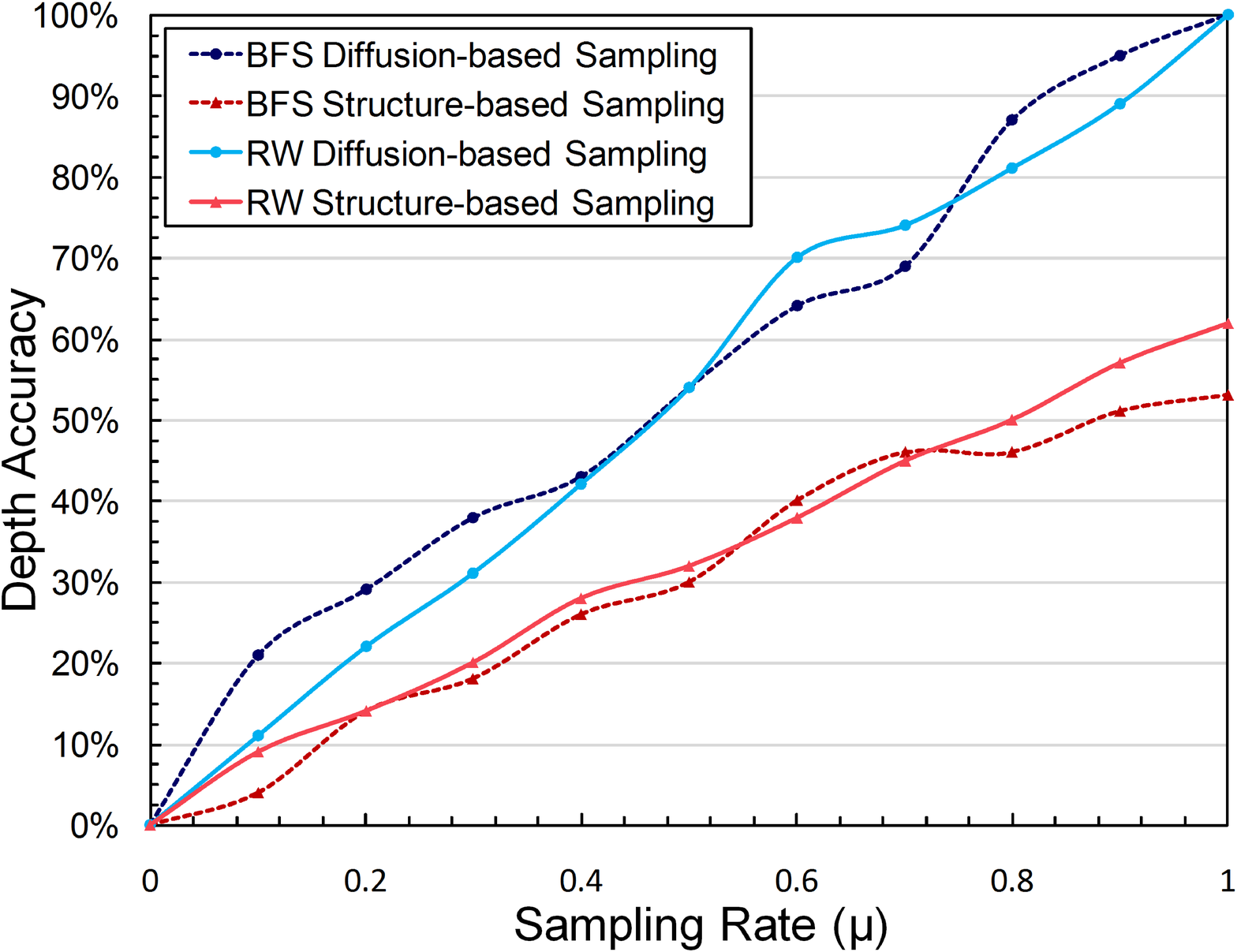}}
    \subfigure[Forest Fire Network]{\includegraphics[scale=0.073]{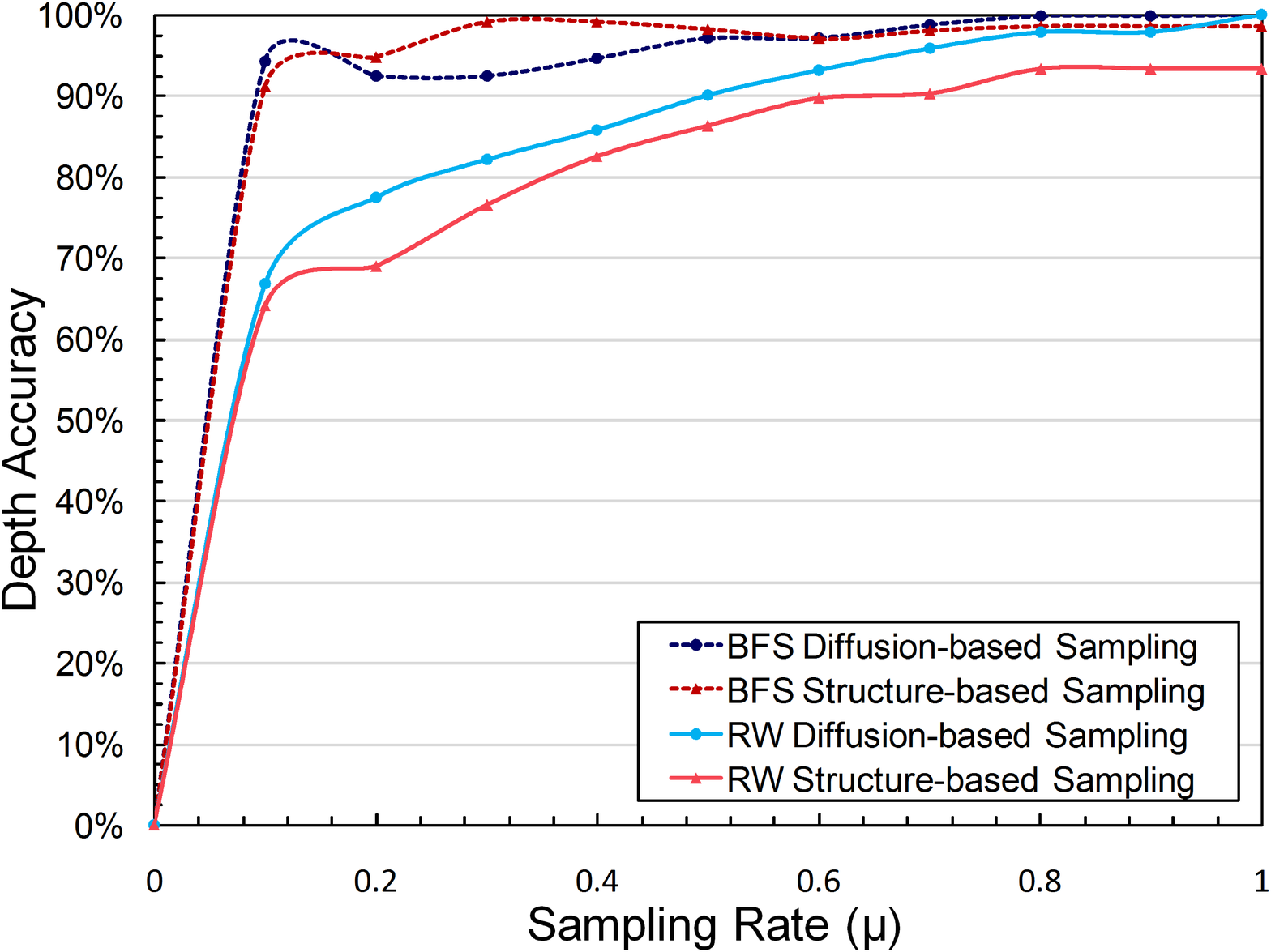}}
    \subfigure[Political Blogsphere Network] {\includegraphics[scale=0.073]{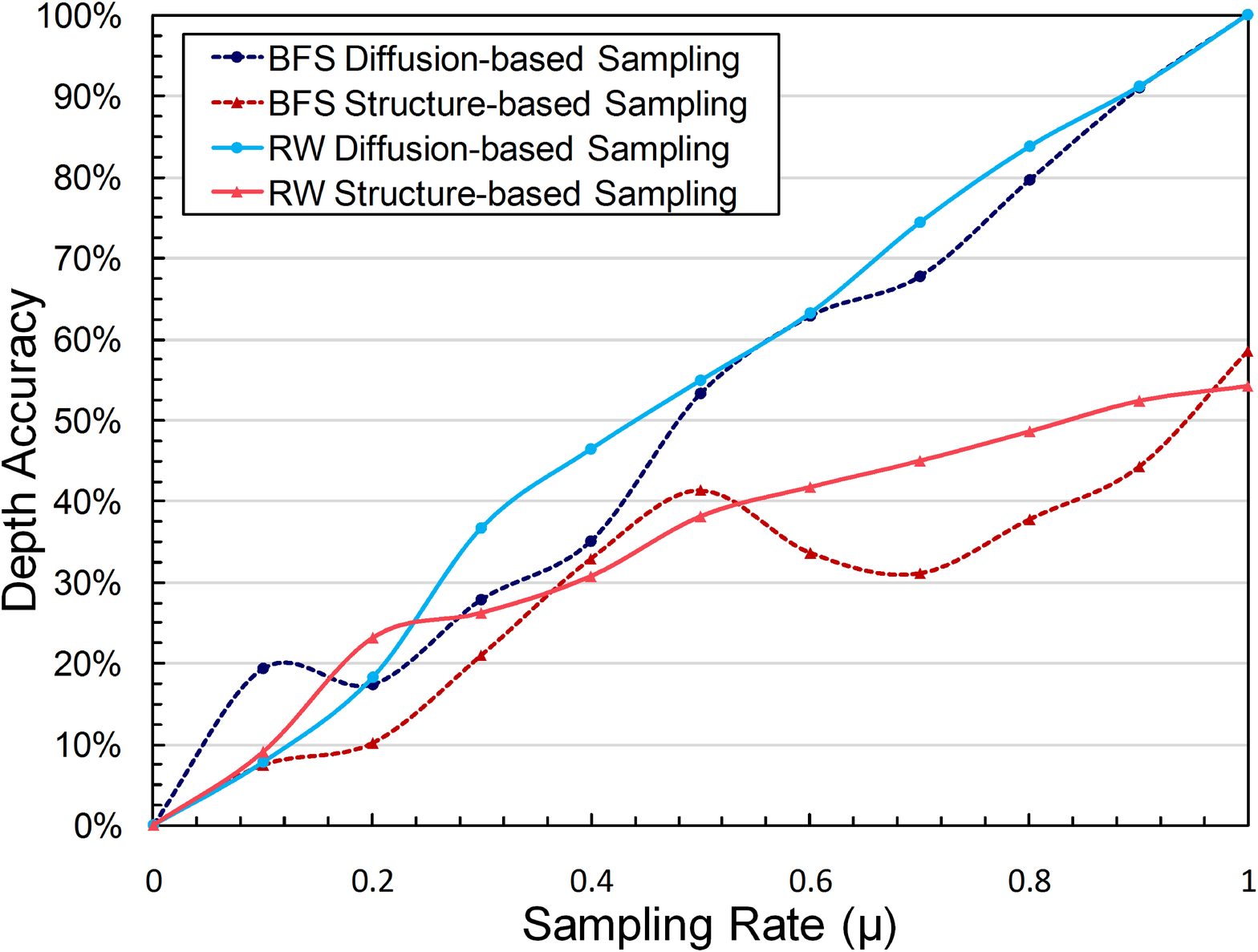}}
     \subfigure[Co-authorship Network]{\includegraphics[scale=0.073]{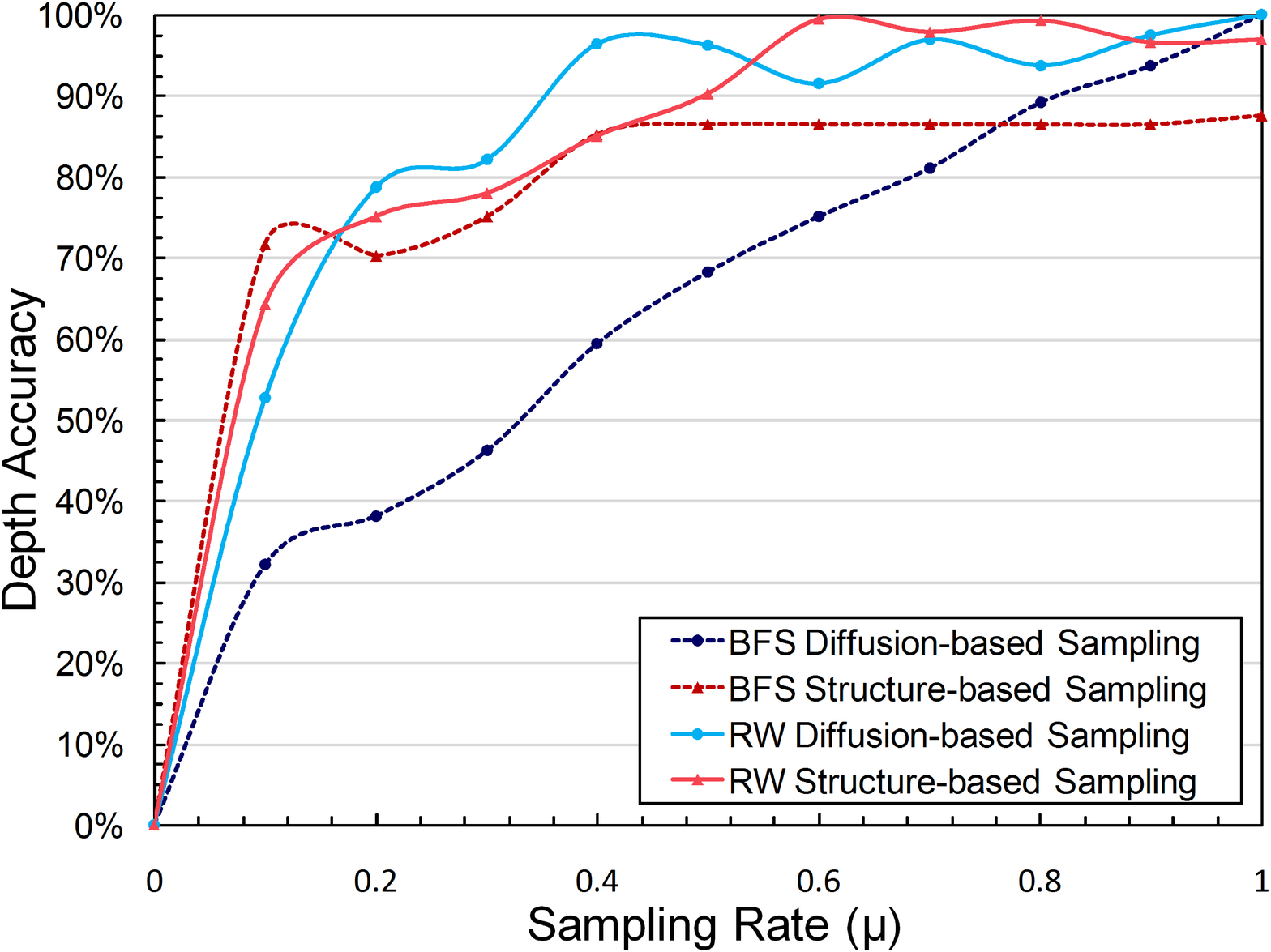}}
  \end{center}
  \caption{\small{Accuracy of depth characteristic measurement in different sampling approaches.}}
  \label{Depth}
\end{figure*}

In previous studies \cite{Nowell08, Choudhury10}, the evaluation of diffusion process is done by measuring some characteristics which are more dependent on the structure of the network rather than its propagation behavior.  In the following, we propose a number of diffusion-based characteristics and classify them into three categories. This classification can be effective in evaluation and analysis of diffusion characteristics. Additionally, the proposed characteristics cover a wide range of measures in gauging diffusion network sampling approaches.

\textbf{Node-based Characteristics.}
The beginners of an infection process play a critical role in the diffusion process. 
In many applications such as political issues, starting a diffusion process is more important than continuing it. Therefore, the beginner of an infection, called ``Seed", can be considered as a node-based characteristic. We define the measurement function $f(u)$ for seed characteristic as follows: $f(u) = 1$, if node $u$ is a seed in the original network, and $f(u) = 0$, otherwise. The previous definitions (e.g. \cite{Choudhury10}), only consider the number of seeds in the sampled network while this new characteristic determines the common seeds between the original and sampled network as a more realistic definition. 

As it was explained in section \ref{sec:related work}, tracking diffusion paths in \textsc{Dbs} approach should result in higher accuracy of analyzing diffusion process in comparison with \textsc{Sbs}. Measuring the accuracy of seed characteristic in both \textsc{Sbs} and \textsc{Dbs} approaches confirms this claim. In Figure \ref{Seed}, the accuracy of seed characteristic measurement has been depicted in all of the synthetic and real networks. Our results show that the seed accuracy in \textsc{Dbs} grows faster than \textsc{Sbs} by increasing the sampling rate. In higher sampling rates, this phenomenon will result in considerable performance difference between these approaches (up to $65\%$).

Nevertheless, \textsc{Sbs} in the blogosphere network can decrease this difference by up to $25\%$ in contrast to the other networks (Figure \ref{blog-seed}). This different behavior is the result of the network density. Considering the relation between the number of nodes and edges given by $E(t) \propto N(t) ^ a$ \cite{LeKlFa05}, the densification exponent ($a$) in the blogosphere network is more than the others (Table \ref{Densification}). This higher density gives \textsc{Sbs} more options in visiting the nodes to find the beginners of the infection. Therefore, \textsc{Sbs} can achieve higher accuracy in a dense network such as blogosphere network.

\begin{table}[h]
\small
\caption{\small{The Networks Densification Exponent.}\label{Densification}} 
\begin{center}
\begin{tabular}[c]{|l|c|c|c|c|c|c|} 
\hline
{ \textbf{Network} } & {\textbf{Dens. Exp. (a)}} \\

\hline
Core-Periphery & $1.06$   \\
\hline
Hierarchical  & $1.03$   \\ 
\hline
Random(ER) & $1.06$    \\
\hline
Forest Fire & $1.03$  \\ 
\hline
Blogshpere & $1.34$ \\
\hline
Co-Authorship & $1.07$ \\
\hline
\end{tabular}
\end{center} 
\end{table}

\textbf{Link-based Characteristics:}
In the diffusion process, some links have more attendance than the others. These links are significant in some applications such as finding potential paths of infection propagation in the epidemic spreading \cite{Eslami11}. Let $C_{e} = \{ c | e\in IV_{c}\}$ be the set of cascades in which link $e$ appears. We define the ``Link Attendance" characteristic by the measurement function $f(e)$ for link $e$ as $f(e) = |C_{e}|$. As shown in Figure \ref{Link}, we can obtain more link attendance accuracy with \textsc{Dbs} compared to \textsc{Sbs}. This performance gap will be greater in higher sampling rates in a manner similar to the seed characteristic.

\textbf{Cascade-based Characteristics:}
In general, the depth of an infection can be determined by the diffusion path length.
Since the diffusion network is usually assumed to be a tree, the depth characteristic is defined by the length of the tree \cite{Nowell08, Choudhury10}. However, a real diffusion network is not a tree. Therefore, we consider the length of a cascade, $c$,  to define the depth characteristic by the measurement function $f(c) = |IV_{c}|$.

As Figure \ref{Depth} shows, \textsc{Sbs} can achieve higher accuracy in depth characteristic compared to the seed and link attendance characteristics. This is the result of inherent difference between these characteristics. More specifically, seed and link attendance are individual-based characteristics while depth is related to the cascades as a group-based characteristic. This feature of the depth characteristic gives \textsc{Sbs} approach more choices in exploring the underlying network. Therefore, the performance difference between \textsc{Sbs} and \textsc{Dbs} will be decreased for depth accuracy from $60\%$ to $35\%$, in average.

\begin{figure}[!b]
\centering
\includegraphics[scale=0.1]{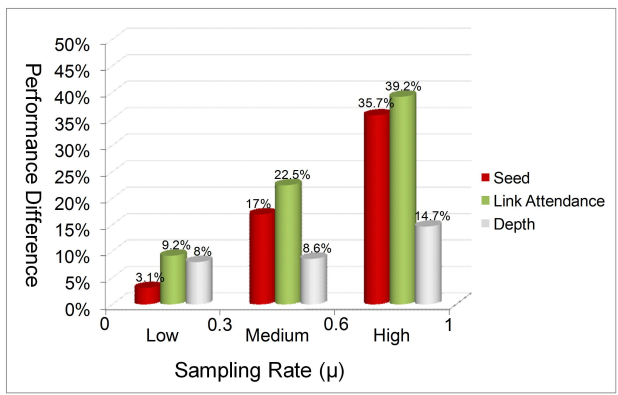}
\caption{\scriptsize{Diffusion-based Sampling (\textsc{Dbs}) vs. Structure-based Sampling (\textsc{Sbs})}}
\label{DS}
\end{figure}

\subsection{Discussion}

Here, we address the general superiority of \textsc{Dbs} vs. \textsc{Sbs} by considering the effect of different sampling rates. The sampling rate has been divided to three ranges; (1) low range: $0<\mu \leqslant 0.3$, (2) medium range: $0.3<\mu \leqslant 0.6$, and (3) high range: $0.6<\mu \leqslant 1$.
Considering these sampling ranges, we first measure the average accuracy of each characteristic for each sampling approach, in all networks. Then we illustrate the superiority of \textsc{Dbs} over \textsc{Sbs} by calculating their performance difference (refer to Figure \ref{DS}). 
Although \textsc{Dbs} performs much better than \textsc{Sbs} in the medium and high sampling ranges (in average by $16\%$ and $29\%$, respectively), the performance difference between them is about $7\%$ in the low sampling rates. Therefore, in real large scale systems that we have to sample the network with a low sampling rate, \textsc{Sbs} would be a better choice because of its lower time complexity in collecting data, compared to \textsc{Dbs}.

\begin{figure}[!h]
\centering
\includegraphics[scale=0.1]{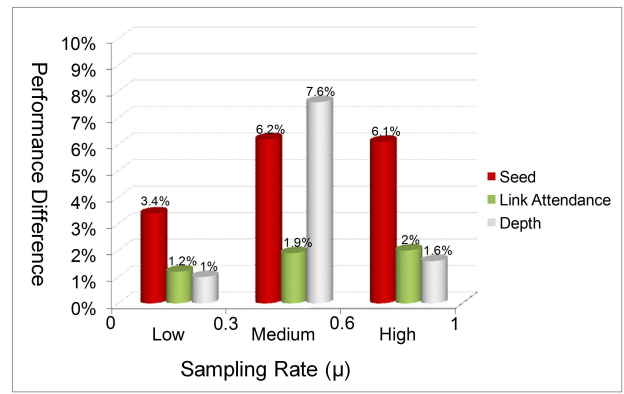}
\caption{\scriptsize{Random Walk (RW) vs. Breadth-First Search (BFS)}}
\label{RB}
\end{figure}

Moreover, we investigated the performance of two sampling methods of RW and BFS for both \textsc{Sbs} and \textsc{Dbs} approaches. Figure \ref{RB} illustrates the superiority of RW with respect to BFS in measuring diffusion characteristics for different sampling rates. However, the performance difference of RW and BFS techniques in average is about $3\%$ which is much lower than the performance difference between \textsc{Sbs} and \textsc{Dbs} approaches. This fact shows the sampling approaches (namely \textsc{Dbs} and \textsc{Sbs}) are more important than the techniques which are used to implement them (namely RW and BFS). 

\section{Conclusions}\label{Conclusions}

In this paper, we introduced the ``Structure-based Sampling (\textsc {Sbs})", and ``Diffusion-based Sampling (\textsc{Dbs})" approaches for the analysis of information diffusion networks. These approaches were evaluated over large synthetic and real networks in terms of the newly proposed diffusion characteristics. Our experiments showed that tracking diffusion paths with \textsc{Dbs} approach will result in more accurate analysis of diffusion process in comparison with \textsc{Sbs}. In addition, by increasing the sampling rates more accurate results are achieved in measuring seed and link attendance characteristics by using \textsc{Dbs}. However, a cascade-based characteristic has different behavior compared to the node-based and link-based characteristics. In this case, the performance difference between \textsc{Sbs} and \textsc{Dbs} in measuring cascade-based characteristics accuracy is decreased.
Furthermore, our analysis on the performance of the introduced sampling approaches showed that structure-based sampling is preferable in the large scale systems in which low sampling rates are more feasible. Moreover, we have found that the sampling techniques such as RW and BFS are less significant than the sampling approaches (i.e. \textsc{Dbs} and \textsc{Sbs}) on analysis of the diffusion process.

We believe that our results provide a promising step towards understanding the sampling approaches in analysis and evaluation of diffusion processes. 
There are several interesting directions for future work. Proposing a new sampling approach which can decrease the gap between structure-based and diffusion-based sampling approaches is one of our main future goals. Including other diffusion aspects such as infection times to define new diffusion characteristics is another aim which we would consider in the future.

\section{Acknowledgments}
This research has been partially supported by ITRC (Iran Telecommunication Research Center) under grant number 6479/500 (90/4/22).

\end{document}